\documentclass[12pt]{iopart}

\usepackage{graphicx}
\usepackage{algorithm2e}
\begin{document}

\title[Static Segmentations in Dynamic PET Images]{Static Segmentations in Dynamic PET Images: The need for a new method}

\author{Laporte, P.$^1, ^2$, Cohalan, C.$^1, ^3$, Carrier, J.F.$^1,^2$}

\address{$^1$Centre de recherche du Centre hospitalier de l'Université de Montréal, Montréal, Québec, Canada\\
$^2$Département de physique, Faculté des arts et des sciences, Université de Montréal, Montréal, Québec, Canada\\
$^3$Department of Physics and Biomedical Engineering, CHUM, Montreal, Quebec H2C0X1, Canada}
\ead{philippe.laporte.3@umontreal.ca}
\vspace{10pt}
\begin{indented}
\item[]August 2022
\end{indented}
\begin{abstract}
	The Task Group 211 report of the American Association of Physicists in Medicine (AAPM) 
	reviewed static segmentation techniques in nuclear positron-emission tomography (PET) imaging used in nuclear medicine. 
	These methods, when applied to a dynamic image, such as the ones obtained in pharmacokinetic analyses, fail to take into account the dynamic nature of the acquisitions.\\
	In this article, the leading hypothesis was that a static segmentation was not adequate in even the simplest dynamic PET images.
	To put this idea forward, a simple dynamic PET phantom was devised. Many dynamic acquisitions were obtained using FDG. \\
	To analyze them, different static segmentations were performed on each timeframe.
	These were followed by quantitative analyses to determine whether the segmentations were consistant between various timeframes of reference. 
	The quantitative analytical tools used were the Sørensen-Dice coefficients, the overlapping of the time-activity curves (TACs),
	and the pharmacokinetic parameters extracted from the images using the Dynesty Python package. 
	In order to perform some of the analyses, an uncertainty had to be added to the TACs themselves: 
	to do so, the individual segmentations were spatially displaced to estimate the sensibility of the TAC to the underlying segmentation.\\
	Using these analytical tools, we propose that static segmentations are not sufficient tools for segmenting dynamic images in a nuclear medicine context. 
	The specific case of pharmacokinetic modelling is used to exemplify this idea. 
	Further work could include a method of estimating uncertainties on segmentations or a novel method for dynamic segmentations in dynamic PET images.
\end{abstract}

\noindent\submitto{Physics in Medicine \& Biology \PMB}
\maketitle
\section{Introduction}
	Nuclear medicine uses radioisotopes to image functional regions of a system. 
	Exactly what is imaged will depend upon the specific drug to which the radioactive atom is attached.
	In the context of pharmacokinetics, it is possible to follow the movement of the radiopharmaceutical through the organism. 
	When coupled with a dynamic acquisition, the exact movement through specific regions can be seen. 
	From this, it is possible to extract relevant parameters, which give information about the metabolic processes relevant to the radiopharmaceutical.
	This metabolic knowledge can allow for more adequate prescriptions of pharmaceutical compounds for patients, 
	which lead to better health outcomes and financial efficiency. With the advent of theranostics, such parameters can play an important role in patient care (Meikle et al. 2021).\\
	To succeed in extracting the relevant parameters from the dynamic positron emission tomography (PET) acquisitions, regions of interest have to be selected. 
	This process can be done manually, by a selection of the voxels specifically or through a shape, or via a more automatic process:
	in both cases, this is a segmentation. 
	Segmentations relevant for static nuclear medicine are highlighted in the Task Group Report 211 of the American Association of Physicists in Medicine (AAPM) (Hatt et al. 2017).
	These static segmentation techniques might not be sufficient in the context of dynamic images, even simple ones. 
	\textbf{This article aims to put forward the inadequacy of static segmentation techniques in the context of quantitative dynamic nuclear medicine acquisitions.}
\section{Theoretical Aspects}
	\subsection{Dynamic PET imaging}
		A dynamic PET acquisition consists of many timeframes, each representing one static PET image. 
		PET images being functional, a dynamic acquisition represents the evolution of a functional system through time (Maguire et al. 2003). 
		These acquisitions can be used to monitor the evolution of a specific functional region and to see how the radiopharmaceutical moves throughout the organism.
		Each timeframe of a dynamic PET acquisition is thus, truly, a static PET acquisition. 
		This implies that each aspect pertaining to the loss of resolution or of noise present in a static PET acquisition 
		will be present in each individual timeframe of a dynamic PET acquisition.
		This leads to the notion that the presence of noise, the choice of reconstruction algorithm, and the specific radioisotope used 
		will influence the quantitative results (Boellaard, 2004, Boudraa et al. 2006).
		For this reason, a dynamic PET acquisition, just like a static one,
		often uses a method of anatomical imaging to correct for scattering and attenuation, such as a computerized tomography (CT) 
		or magnetic resonance imaging (MRI) image.\\
		In a dynamic PET setup, additional analytical difficulties are present, the most prevalent one being the presence of movement between timeframes. \\
		For a static acquisition, movement during the imaging time will lead to a blurring of the image and errors in the coregistration of the anatomical image, 
		caused by the displacement of the emitting source (Kirov et al. 2012):
		in a dynamic context, this type of movement will always be intra-timeframe (Cherry and Dahlbom 2006), i.e. during a specific timeframe.
		In the dynamic context, there can also be movement between the timeframes, i.e. inter-timeframe movement.
		This type of movement will cause a displacement of the functional regions. 
		Being a functional means of imaging, nuclear medicine is ill-suited to discriminate between the movements 
		caused by the metabolization of the radiopharmaceutical and the movements caused by a displacement of the body itself.
		In the latter category are included the filling of a bladder or the twitching of the body.
		For these types of movement to be of analytical relevance in a dynamic setting, they have to either be non-periodic or with a period greater than the length of a timeframe;
		otherwise, they will fall within the blurring aspect of a static PET acquisition.\\
		This non-periodic movement between timeframes is what is hypothesized will play a major role in the insufficiency of the actual 
		static segmentation schemes in the context of dynamic quantitative nuclear medicine.
	\subsection{Segmentation}\label{segmentationSec}
		In this paper, region of interest (ROI) selection is done via a segmentation, from which a time-activity curve is extracted.
		It allows to create a new binary image from an initial image, where one label represents the region of interest (0's and 1's).
		A segmentation can be done either manually or (semi-) automatically. \\
		Manual segmentations are prone to many errors, such as inter- and intra-user variability (Fahey et al. 2010, Hofheinz et al. 2012). 
		The experience of the user, his level of energy and awareness, the conditions of the room where the user is located, and many other factors 
		have a significant impact on the obtained results.
		The specific parameters of the acquisition also impact the resulting delineation of a region of interest (Hatt et al. 2011).
		Even in the case of static PET images, the process of selecting an ROI has an impact on the quantitative results (Boellaard, 2009).\\
		For this reason, the authors of this paper have opted for automatic segmentation schemes, 
		as put forward by the Task Group 211 of the American Association of Physicists in Medicine.
		This allows for a higher level of reproducibility, which is an important goal of the scientific endeavour (Lee, 2010).
		The TG-211 presents genres of segmentation methods for static nuclear medicine images (Hatt et al. 2017). 
		The most prominent are those based upon gradients, those based upon statistics and those based upon filling methods.
		Schemes based upon machine learning were not evaluated in the present article. 
		\subsubsection{Gradient-Based Methods}
			Methods based on the gradients of the image use the fact that contours are represented by a quick variation in the intensity of neighbouring voxels.
			The difference in intensity of two spatially adjacent voxels in a 1-dimensional image is the definition of the derivative limited on a discrete domain along a line,
			\begin{equation}
				I' = I(x) - I(x-1) = \lim_{h\rightarrow 0^+}\frac{I(x+h)-I(x)}{h},
				\label{gradient}
			\end{equation}
			where $I(x)$ is the intensity at a location $x$.\\
			By specifying a lower threshold on the value of the derivative, it is possible to determine the contour of a region of interest.
			The same principle can be extended to a 3-dimensional image, 
			by looking at the norm of the gradient of the three spatial dimensions, i.e. $\Vert \nabla I\Vert $.\\		
			In the present study, the method used is based on the Canny edge detection algorithm, as represented by the algorithm \ref{alggrad} (Canny, 1986). 
			This algorithm is applied by taking a subsection containing only the region of interest and using the algorithm on this subregion.\\
			\RestyleAlgo{boxruled}
			\begin{algorithm}[ht]
				\label{alggrad}
				\caption{Gradient-Based segmentation (Canny method)}
				1.0. Determine the 2D contours for all slices along each axis independently:\\
				1.a. \qquad Apply a Gaussian filter;\\ 
				1.b. \qquad Compute a gradient norm and gradient direction images;\\
				1.c. \qquad Select only the pixel whose value is maximal along the orientation of the gradient;\\
				1.d. \qquad Threshold the values by hysteresis: all pixels above $\tau_{high}$ are part of the contour and 
				those above $\tau_{low}<\tau_{high}$ are part of the contour if and only if they are adjacent to a contour. 
				$\tau_{high}$ and $\tau_{low}$ are the values of the upper and lower thresholds, respectively;\\
				(1.e.) \qquad This leads to a 2D contour of the region of interest, contained on a slice of the image.\\
				2.0. For each 2D contour, fill it:\\
				2.a. \qquad Determine the center of mass of the contour;\\
				2.b. \qquad From each voxel of the contour, iteratively go towards the center of mass;\\
				(2.c.) \qquad This leads to a 2D region of interest, contained on a slice of the image.\\
				3.0. For the 3D image, a voxel is part of the region of interest only if it is part of the region of interest on the corresponding 2D slice for all axes.
			\end{algorithm}
			This method takes into account a few premisses, mainly that the image has been well limited around the region of interest, 
			that the shape has a center of mass within its contour and that it is convex. 
			These assumptions are here justified, since the region of interest will be a saline pouch, i.e. a closed and roughly ellipsoid shape.
		\subsubsection{Statistics-Based Methods}
			Statistics-based segmentation methods use the underlying assumption that the different classes of interest in an image belong to underlying 
			statistical distributions. 
			The idea is to try to assign each voxel to the distribution to which it should statistically belong (Hatt et al. 2017).\\
			In the present study, it is assumed that the region of interest and the rest of the image are represented by two normal random variables. 
			The exact algorithm used is based on an Iterative Conditional Modes algorithm (ICM), as introduced in algorithm \ref{algstat} (Besag 1986). 
			This algorithm adds a contributing factor to consider the class to which belong neighbouring voxels:
			this is to take into account the notion that spatially close voxels have a higher chance of belonging to the same underlying random variable.
			\begin{algorithm}[ht]
				\label{algstat}
				\caption{Statistics-Based segmentation (ICM)}
				1. Select the highest and lowest values of intensity as $\mu_1$ and $\mu_2$, where $\mu_i$ represents the mean of the $i^{th}$ class; \\
				2. Set $\sigma_1=\sigma_2=(\mu_2-\mu_1)/2$, where $\sigma_i$ represents the standard deviation of the $i^{th}$ class;\\
				3. Iteratively, until stability of the segmented regions between two iterations:\\
				4. \qquad compute $\mu_1$, $\mu_2$, $\sigma_1$, and $\sigma_2$ from the two regions;\\
				5. \qquad for each voxel of the image with intensity $I$:\\
				6. \qquad\qquad Check to which normal distribution it belongs, i.e. compute
				$$\ln(\mathcal{N}(I;\mu_1,\sigma_1))\textnormal{ and } \ln(\mathcal{N}(I;\mu_2,\sigma_2)),$$
				where $\mathcal{N}(I;\mu_i,\sigma_i )$ represents the probability to obtain 
				the value $I$ from a normal distribution with mean $mu_i$ and standard deviation $\sigma_i$.  \\
				7. \qquad\qquad Check the number of neighbours $n_{1,2}$ of classes 1 or 2\\
				8. \qquad\qquad Check which is higher: 
				$$\ln(\mathcal{N}(I;\mu_1,\sigma_1))+\alpha n_2 \textnormal{ or } \ln(\mathcal{N}(I;\mu_2,\sigma_2))+\alpha n_1,$$
				where $\alpha$ is a hyperparameter given by the user.\\
				9. \qquad\qquad Set the voxel to the class to which it probabilistically belong\\
			\end{algorithm}
		\subsubsection{Filling-Based Methods}
			Filling algorithms start from a given seed of voxels and expand, following conditions on whether to incorporate any voxel to the region of interest.
			Two basic conditions for the incorporation of a voxel are for it to be adjacent to the current volume of interest and for its intensity to be within a certain range,
			normally given by the mean $\mu$ and standard deviation $\sigma$ of the currently segmented region (Adams et al. 1994). \\
			In the present case, the range is given by $\left[\mu-f\sigma,\mu+f\sigma\right]$, where $f$ is a hyperparameter given by the user. 
			Since this parameter is highly dependent upon the image and the initial condition, a recommended approach is to segment for an array of factors $f$
			and to compare the volume segmented with respect to the values of $f$. 
			The optimal $f$ for a segmentation is then set to be the one just before the volume increases significantly, since it represents the moment when the filling 
			"pours" out of the region of interest to fill the background (Hatt et al. 2017). \\
			This model also runs on a set of premisses, mainly that the shape is spatially closed, so that the filling is not smoothly gradual and so that there will be a
			moment of steep increase for the volume. 
			\begin{algorithm}[ht]
				\label{algfill}
				\caption{Filling-Based segmentation}
				1. Start with a seed and a range of factors $f$;\\
				2. For a given factor $f$:\\
				3. \qquad Iteratively, until stability of the segmented region between two iterations:\\
				4. \qquad\qquad Compute the mean $\mu$ and the standard deviation $\sigma$ of the current region;\\
				5. \qquad\qquad For each voxel adjacent to at least one voxel of the region, add it to the ROI if its intensity is between 
				$$[\mu-f\sigma,\mu+f\sigma];$$\\
				(5.) \qquad\qquad This gives a segmented volume by filling for a given value of $f$;\\
				6. The segmentation kept is the one for which the total segmented volume increases by a certain factor, say 150\%, for the next value of $f$.
			\end{algorithm}
	\subsection{Pharmacokinetic Model}
\begin{figure}
			\includegraphics[width=\linewidth]{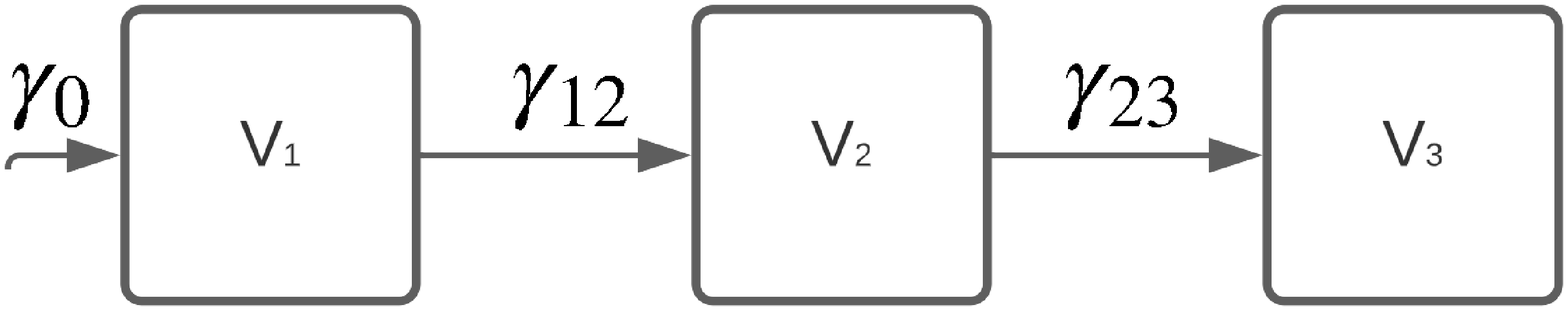}
			\caption{Two-compartment model for a pharmacokinetic process.}
			\label{model}
		\end{figure}

		A pharmacokinetic model is based on a set of assumptions about the movement of the radiopharmaceutical through a given dynamic system.
		Here, the adopted model is a two-compartment model, where the radiotracer is entirely in the first compartment at time 0 (Maguire et al. 2003). 
		There is a constant flow of liquid devoid of radioactivity into the first compartment, forcing a flow into a second one, and then into a third one.
		Figure \ref{model} illustrates the schematic of the model.
		\\This model is governed by the following set of differential equations:
		\begin{equation}
			\centering
				\pmatrix{
					Q'_1 \cr Q'_2 \cr Q'_3
				}
				=
				\pmatrix{
					-\frac{\gamma_{12}}{V_1} & 0 & 0\cr
					\frac{\gamma_{12}}{V_1} & -\frac{\gamma_{23}}{V_2} & 0\cr
					0 & \frac{\gamma_{23}}{V_3(t)}& 0
				}
				\pmatrix{
					Q_1 \cr Q_2 \cr Q_3
				}
				\label{diff}
		\end{equation}
		In equation \ref{diff}, the $\gamma$'s are the transfer coefficients, the $V$'s are the volumes of the compartments, 
		the $Q$'s are the quantity of radiotracer, and the derivatives are with respect to time.\\
		Some of the underlying assumptions of such a model are the following:
		\begin{enumerate}
			\item The mixing of the radiopharmaceutical in the compartment is instantaneous. 
				As soon as a quantity of radiotracer enter the compartment, it goes uniformly everywhere and does not stick to the walls of the compartment;
			\item The coefficients of transfer are constant through time, i.e. $\gamma_{ij}' = 0$;
			\item There is no reverse flow of radiotracer from a compartment to the previous one. In other word, there is a $\gamma_{12}$, but $\gamma_{21} = 0$.
		\end{enumerate}
		With this model and these assumptions, this linear system of differential equations can be solved for the first two compartments (1 \& 2). \\
		In this case:
		\begin{equation}
			Q_1(t) = \frac{Q_0}{V_1}e^{-\frac{\gamma_{12}}{V_1}t}
			\label{Q1}
		\end{equation}
		\begin{equation}
			Q_2(t) = Q_0\left(\frac{\gamma_{12}}{\gamma_{23}V_1-\gamma_{12}V_2}\right)\left[e^{-\frac{\gamma_{12}}{V_1}t}-e^{-\frac{\gamma_{23}}{V_2}t}\right]
			\label{Q2}
		\end{equation}
		By segmenting a region of interest, it will be possible to extract a time-activity curve, which will monitor the quantity of radiopharmaceutical in the compartments.
		It will then be possible to fit these curves with equation \ref{Q2} to obtain the pharmacokinetic parameters.
\section{Experimental Setup}
	The machine used for the acquisitions was a Siemens Biograph mCT Flow, with the software PETSyngo62C. 
	For each acquisition, 40 timeframes of 60 seconds were obtained, 
	for a total acquisition time of 40 minutes. A CT scan was done before the PET acquisition.
	The vendor's OSEM reconstruction with time-of-flight and point spread function (TOF PSF) was used. 
	Corrections were applied for random and scatter detections, dead time, and attenuation. \\
	For the acquisitions, a custom-made phantom was devised. 
	It consisted of a small vial and two water pouches, which were linked, as in figure \ref{model}, sequentially with syringes and catheters. 
	An initial water pouch was used to induce a water flow by gravity, starting in the vial.
	The following compartments were placed spatially lower in sequence, at different heights.
	All compartments were filled to full capacity with tap water, except for the last water pouch, where all the liquid would go. \\
	The water flow into the system was started before the start of the PET acquisition. 
	After starting the acquisition, 32.9 to 60 MBq of FDG were injected into the vial within 20 seconds, 
	forcing the FDG to mix with the water in the vial and flow out into the next two compartments.
	In total, 7 phantoms were imaged. Out of these, 2 acquisitions were problematic for experimental reasons (leakage of the fluids, unplugging during imaging, etc.). 
	The total number of valid acquisitions was thus n = 5. 
\section{Numerical Analyses}
	The analyses were done using Python 3.8.13, with specialized packages, including Dynesty (Speagle et al. 2020), PyDicom (Mason et al. 2022), 
	Numpy, Scipy and a inhouse code.
	For each acquisition, a subsection of the whole acquired image was selected around the first water pouch. 
	This represented compartment 2 of the model.
	The selected subsection was done such that the compartment of interest was always within the subsection, for all timeframes; this was confirmed visually.
	Each of the three methods of static segmentation was then applied to each subsection for every timeframe. 
	A timeframe of reference will be defined as a timeframe from which segmentations will be produced, notwithstanding the other timeframes.
	A visual analysis of the segmented region was then done: all those which looked roughly like the desired volume of interest were kept; the others were discarded.\\
	From the remaining segmented volumes, three tools were used to analyze their similarities: 
	Sørensen-Dice coefficients, the time-activity curves, and the pharmacokinetic parameters.\\
	The first compartment, i.e. the vial, was not used for the analyses, due to its small size and the presence of unmoving radiotracer in the syringe.
	\subsection{Sørensen-Dice Coefficients}
		The Sørensen-Dice coefficient is used to compare the similarity of two sets (Hatt et al. 2017).\\
		For sets $A$ and $B$, the coefficient is given by
		\begin{equation}
			D(A,B) = \frac{2|A\cap B|}{|A| + |B|},
			\label{Dice}
		\end{equation}
		where $|\cdot|$ represents the cardinality, i.e. the size, of the set.\\
		A Dice coefficient of 1 represents a perfect overlap between the sets, whereas a value of 0 means that there is no overlap between them.\\
		For a given dynamic acquisition, a given segmentation method, and for any 2 segmentations, the Dice coefficient was computed.
		This allowed to see how similar the obtained segmentations were. 
	\subsection{Time-Activity Curves (TACs)}
		To compare the TACs, a small methodological error was introduced in each valid segmentation. 
		To implement it, each segmentation was shifted along the three spatial axes by one voxel (+1 \& -1). 
		This one voxel shift is to estimate a minimal sensibility for a binary segmentation of the image.
		This method of shifting created 6 new segmentation volumes and 6 accompanying TACs.
		These seven TACs (original + 6 shifted) gave a new averaged TAC and their standard deviation was the error associated with it.\\
		By doing such a thing, it was possible to see whether the TACs would be overlapping in their uncertainty and statistically similar.
	\subsection{Pharmacokinetic Parameters}
		Using the TACs with uncertainty from the last subsection and the analytic model from equation \ref{Q2},
		the pharmacokinetic parameters were extracted for the compartment. 
		The tool used was the Python package \textit{Dynesty} (Speagle et al. 2020), which uses Bayesian statistics to estimate the distribution of the parameters. 
		In this specific case, the method used is nested sampling to determine the posterior distributions (Higson et al. 2017, Skilling 2006).
		This yields an estimated value and its associated range. 
		For coefficients statistically similar, they would have to fall within each other's uncertainties.\\
		To reduce the computational complexity, the parameters extracted were $Q_0\left(\frac{\gamma_{12}}{\gamma_{23}V_1-\gamma_{12}V_2}\right)$, $\gamma_{12}/V_1$, and $\gamma_{23}/V_2$.
\section{Results}
\begin{figure}
		\centering
		\includegraphics[width=\textwidth,height=\textheight,keepaspectratio]{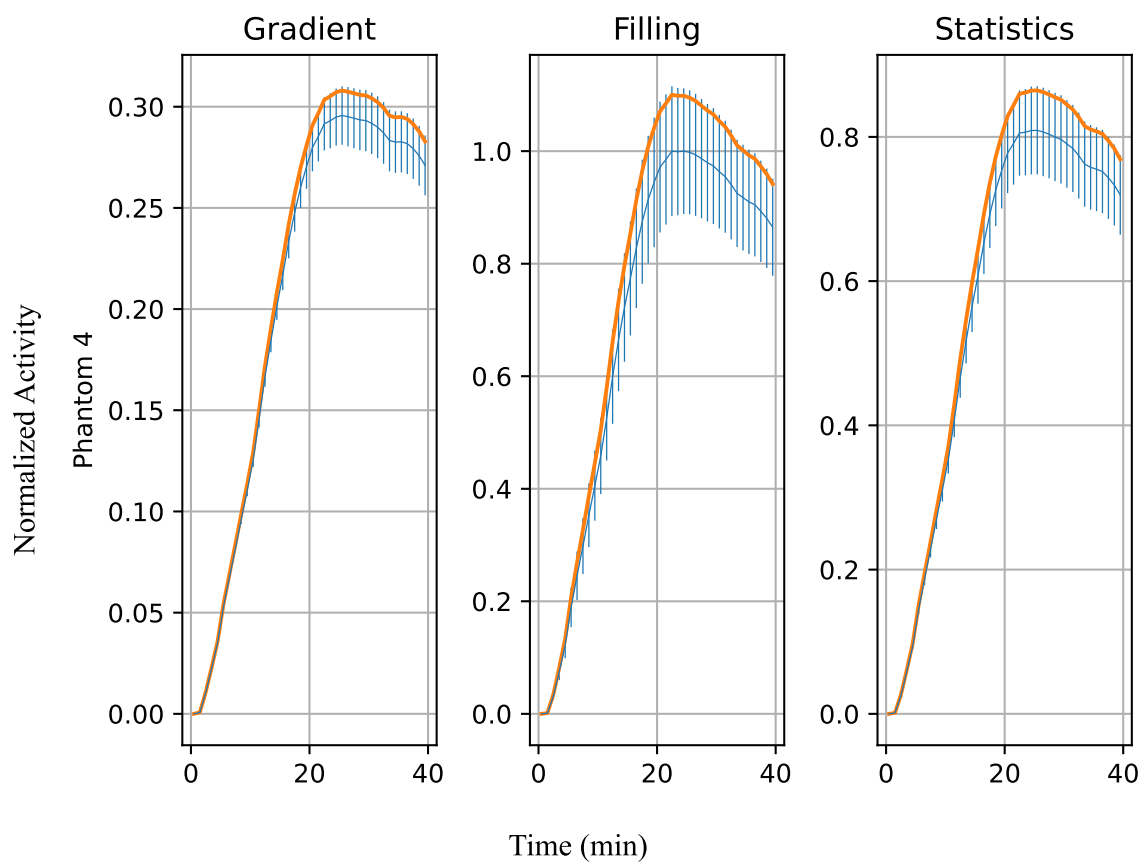}
		\caption{A specific TAC (red) and its linked TAC with errors (blue) for one of the phantoms. 
		The introduction of errors reduce the average value, but the TAC falls within one standard deviation of the error bars.}
		\label{Curves}
	\end{figure}

\begin{figure}
		\centering
		\includegraphics[width=\textwidth,height=\textheight,keepaspectratio]{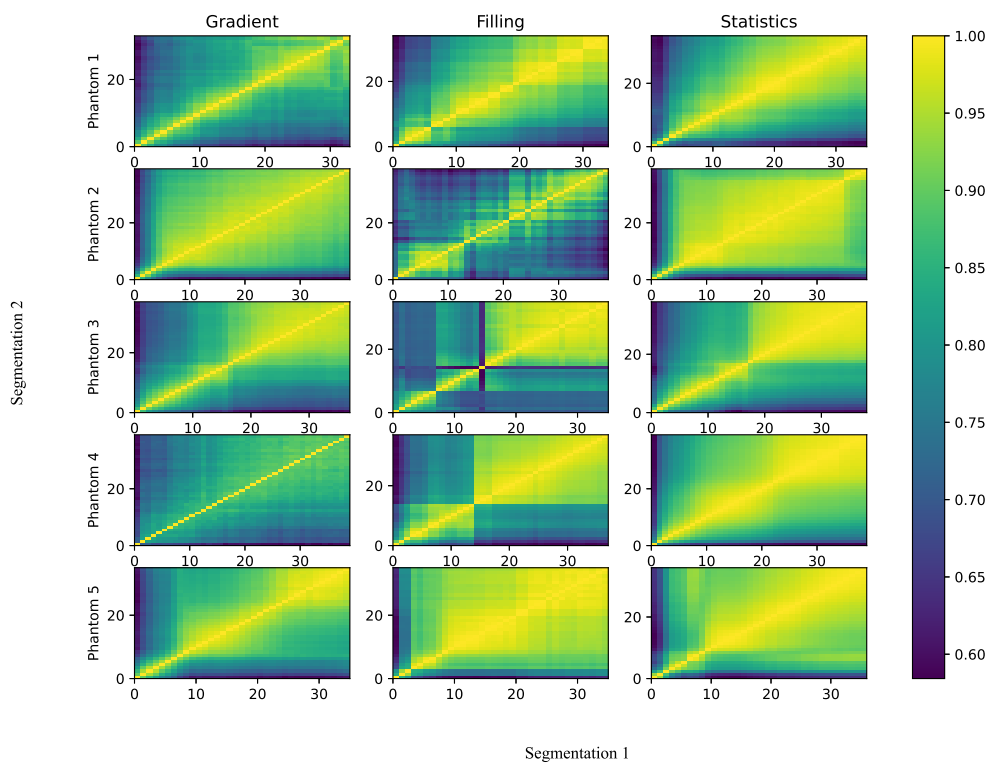}
		\caption{Sørensen-Dice coefficients for the second compartment of the various phantoms. 
		Each subgraph represents a single segmentation scheme for a given phantom. 
		For each subgraph, the axes represent two different timeframes of reference upon which the segmentations are based.}
		\label{Dices}
	\end{figure}

\begin{figure}
		\centering
		\includegraphics[width=\textwidth,height=\textheight,keepaspectratio]{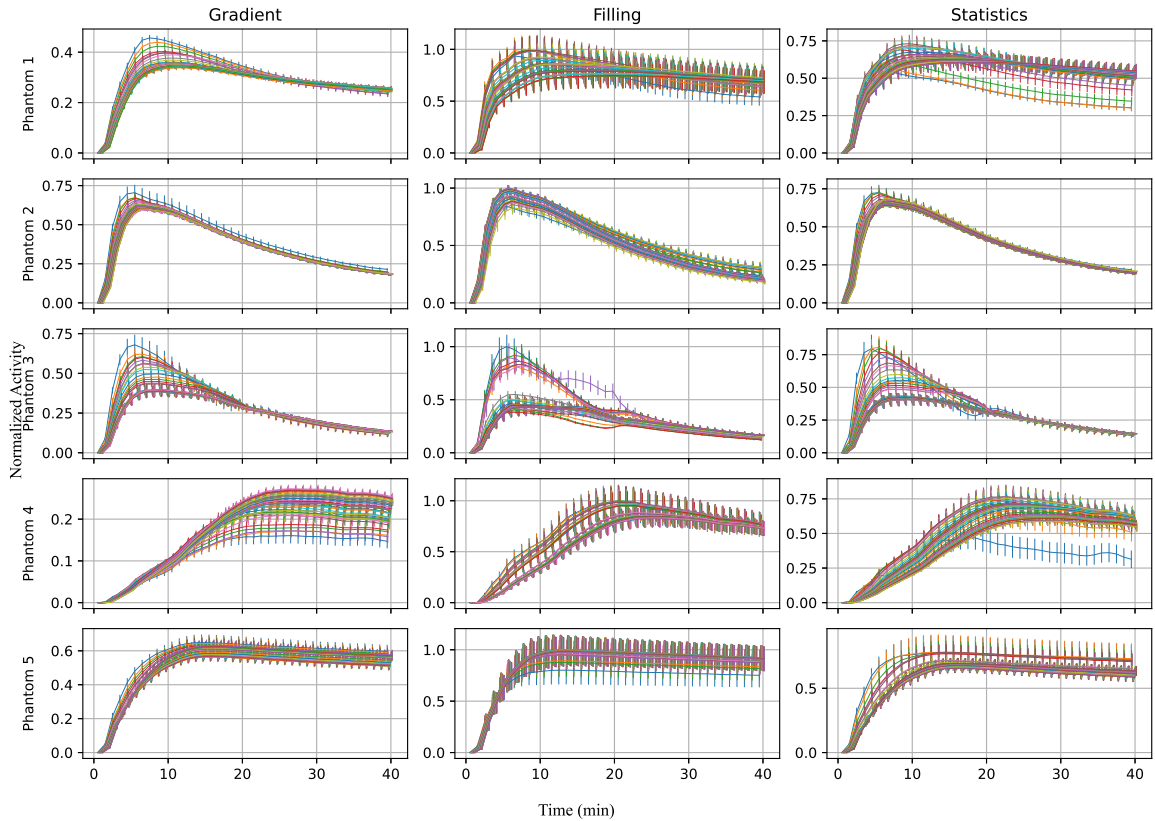}
		\caption{Time-activity curves for the second compartment of the various phantoms. 
		Each subgraph represents a single segmentation scheme for a given phantom; 
		each curve represents a segmentation based on a specific timeframe.}
		\label{TAC}
	\end{figure}

\begin{figure}[h!]
		\centering
		\includegraphics[width=\textwidth,height=\textheight,keepaspectratio]{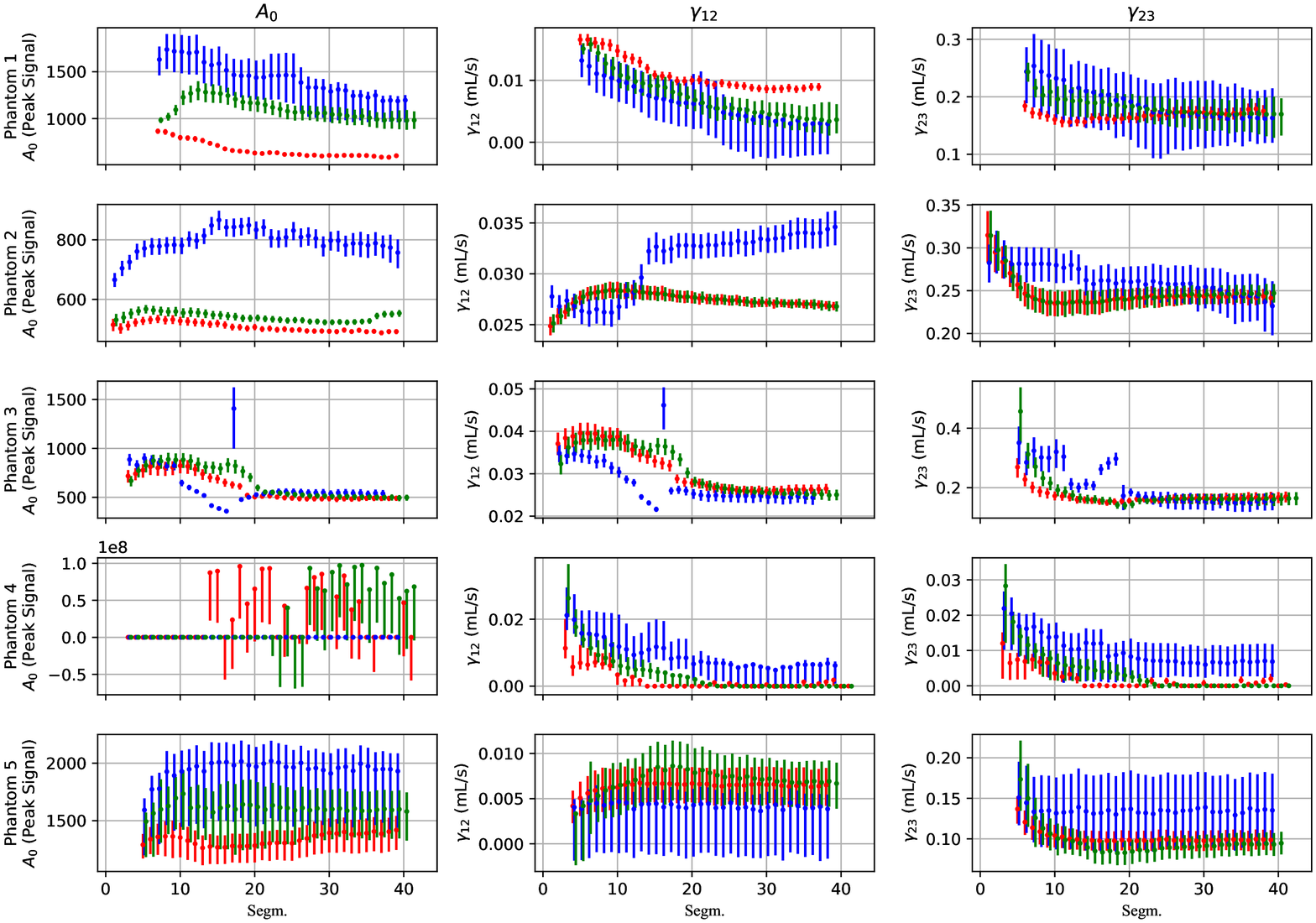}
		\caption{Pharmacokinetic parameters extracted for the different phantoms. 
		Each subgraph represents one parameter. The red, blue, and green lines are respectively for the gradient, filling, and statistics segmentations.
		The x-axis represents the timeframe upon which a given segmentation has been based.}
		\label{Dynesty}
	\end{figure}

	For each of the 5 acquisitions, the previously introduced numerical analyses were conducted. 
	Figure \ref{TAC} illustrates the resulting TACs and TACs with errors for the three segmentation schemes for a specific timeframe of reference on a specific phantom.
	The Dice coefficients have a low level of similarity and have a strong correlation along the diagonal axis, 
	i.e. when the two compared timeframes of reference are close in time, as represented in figure \ref{Dices}. 
	Their value varies, for most cases, between 0.4 and 1.0. 
	This indicates a low degree of similarity between the various segmentations. 
	In other words, the results vary significantly based on the timeframe used to perform the segmentation.\\
	The TACs with a small methodological error show a higher degree of similarity, 
	since, even with just a single voxel shift, they overlap in their respective uncertainties, as illustrated by figure \ref{TAC}.
	A noteworthy fact to point out is that, in general, the central value of the TAC with an error is slightly lower than the TAC without the error, as illustrated in figure \ref{Curves}. 
	This is indicative of the fact that our 1 voxel displacement to estimate error adds a certain blur to our ROIs.\\
	The pharmacokinetic parameters do not show an overlapping in their estimated values, as expressed by their degree of variability. 
	This can be seen in figure \ref{Dynesty}.
	Although for two successive timeframes they do overlap, it is not necessarily the case for segmentations based on more distant timeframes.\\
\section{Discussion}
	This experimental setup represents one of the simplest dynamic model possible. 
	The premise was that for a static segmentation to be valid on a dynamic acquisition, it would have to be sufficient on the simplest of case.
	In this study, it would have been expected that all three quantitative tools yield similar and concordant results, i.e. that the segmentations be valid.
	This is the result that can be inferred by looking at the TACs with errors (figure \ref{TAC}), since there is a lot of overlap.\\
	The Dice coefficients (figure \ref{Dices}) indicate however that the segmentations are not as similar as the TACs lead us to believe. 
	This numerical result is significant, since it is difficult to visually see actual differences in the segmented volumes of interest between two subsequent timeframes. 
	The variations being small and hard to see implies that two segmentations done manually would have been highly similar. 
	The variation between timeframes is quite small, yet the actual Dice coefficients indicate a good variation for the exact volume of interest. 
	This variation in Dice coefficients might actually be caused by liminal variations, i.e. coming from the periphery of the volume of interest,
	where there is often a partial volume effect due to the physical dimensions of the crystals of the detecting device (Soret et al. 2007).
	Physically, this could be explained as coming from the mixing of the radiopharmaceutical.
	As had been put forth in the model of the phantom, it was taken for granted that the FDG would go all around the compartment instantaneously. 
	In reality, this is not exactly the case:
	although the pouch was filled fully with water, 
	it takes time for the FDG to reach the outermost peripheral zones of the compartment and
	a small variation of the volume can happen due to unequal transfer of liquid within the compartment and stretching of the pouch.\\
	For the pharmacokinetic parameters extracted, they varied depending upon which timeframe was used as the basis of the segmentation, as seen in figure \ref{Dynesty}.
	These variations were quite significant, since they did not agree between segmentation schemes. 
	Furthermore, they did not even agree within a single segmentation process and varied quite considerably, especially for distant timeframes.
	Many reasons can explain this phenomenon: 
	when the model was put forth, a basic assumption was that there would be an instantaneous and perfect mixing of the FDG, which is not the case.
	To see it more clearly, the FDG was mixed with a colouring substance, which dyed the tap water when the mixing occured. 
	The dyed FDG was not always mixing perfectly and had a tendency to stick close to were it came in and at the bottom of the compartment. 
	If the initial premises are not satisfied, it is subsequent that the analytic model not fit perfectly. 
	However, the extracted parameters should still be relatively similar, if the segmentations were to be considered equivalent, since they are forced upon the same model.
	This leads to the notion that even though the time-activity curves might appear similar, 
	the actual similarity between the segmented volumes and the extracted parameters are much lower. 
	A visual assessment of the accuracy of the segmentations might not be sufficient, which was an original reason to undergo this study.\\
	Even more, a direct assessment of a single segmentation based on a single timeframe can miss the wide range of the values of the 
	parameters that could have been extracted.
	Consider the phantom 2 $\gamma_{12}$ subgraph of figure \ref{Dynesty}. 
	Even for a single segmentation algorithm, the computed parameters can vary by a one order of magnitude, which is substantial 
	(one can take the $\gamma_{12}$ parameter for phantom 3).  
	The difference between the highest and the lowest possible bounds is much higher than any single uncertainty on a given value. 
	Had a single static segmentation been done, it could be possible to underestimate the range of the parameters by a large margin.
	It thus makes sense to not limit a numerical analysis of dynamic images to a single segmentation based on a static timeframe.\\
	As mentioned in section \ref{segmentationSec} machine learning algorithms have been neglected. 
	The reason is that it would be difficult to evaluate the accuracy and precision of the results.
	Part of the reason is that the ground value of the pharmacokinetic parameters is not known: 
	it would thus be problematic to train a model to determine the optimal segmentation. 
	Besides, the goal of this present paper was to raise the difficulty of static segmentations in a dynamic context. 
	Using a Popperian vision (Popper 2002), the goal is to determine a condition to disprove the adequacy of current methods, which the proposed 3 methods satisfy.
	The approach of counter-factual scenarios (Foley 2012) leads to a similar principle: the interest lies in the ability to create counter-factual scenarios, 
	which here can happen by the ability of the user to modify the hyperparameters manually, whereas this is done by the machine learning algorithm by itself.

\section{Conclusion}
	Segmentations as introduced by the TG-211 of AAPM fall short in the context of dynamic acquisitions.
	Even in the simplest case, i.e. FDG going through static compartments of a phantom based solely on gravity,
	the methods yield hugely varying results depending on which timeframe is used as reference. 
	This impact can be seen mainly through the Sørensen-Dice coefficients and the extracted pharmacokinetic parameters.
	It is thus the authors' opinion that new methods should be put forth that take into account 
	the various aspects of dynamic acquisitions making them differ significantly from static acquisitions.
	Another point of further study would be concerning the use of margins of errors on a single time-activity curve.
	A segmentation algorithm that could also lead to a time-activity curve with an uncertainty could circumvent some of the problems highlighted in this article,
	as it would concede that the segmentation methods are not methodologically statistically perfect.
	This could be an alternative route of enquiry compared to finding the most exact segmentation scheme for dynamic acquisitions.
	In all cases, the next steps seem to be to make the segmentation schemes in a dynamic setting more reliable, reproducible, and transparent.
\ack
	This research was conducted as part of the activities of the TransMedTech Institute, 
	thanks in part to the financial support of the Apogee Canada Research Excellence Fund and the Fonds de Recherche du Québec.
\section*{References}
\begin{harvard}
	\item[] Adams, R, Bischof, L. 1994. Seeded Region Growing. \textit{IEEE Transactions on Pattern Analysis and Machine Intelligence}. \textbf{16(6)}: 641-7.
	\item[] Besag, J. On the Statistical Analysis of Dirty Pictures. 1986 \textbf{48(3)}:259-79
	\item[] Bredies, K and Lorenz, D. 2018. \textit{Mathematical Image Processing}. Cham, Switzerland: Birkhäuser.
	\item[] Boellaard, R Krak, N C , Hoekstra, O S and Lammertsma, A A  2004. Effects of Noise, Image Resolution, and ROI Definition on the Accuracy of Standard Uptake Values: A Simulation Study. \textit{Journal of Nuclear Medicine}. \textbf{45}:1519-27.
	\item[] Boellaard, R. 2009. Standards for PET image acquisition and quantitative data analysis. \textit{Journal of Nuclear Medicine}. \textbf{50}:11S-20S.
	\item[] Boudraa, A O and Zaidi, H  2006. 'Image segmentation techniques in Nuclear Medicine Imaging', in Zaidi, H. (ed.) Quantitative Analysis in Nuclear Medicine Imaging. London, UK: Springer Verlag.
	\item[] Canny, J. 1986. A Computational Approach to Edge Detection. \textit{IEEE Transaction on Pattern Analysis and Machine Intelligence}. \textbf{8(6)}:679-98.
	\item[] Cherry, S R and Dahlbom M. 2006. \textit{PET: Physics, Instrumentation, and Scanners}. New York, NY: Springer Science+Business Media.
	\item[] Diaz, A M A, Drumeva, G O, Laporte, P, Alonso Martinez, L M, Petrenyov, D R, Carrier, J F et al. 2021. Evaluation of the high affinity [$^{18}$F]fluoropyridine-candesartan in rats for PET imaging of renal AT$_1$ receptors. \textit{Nuclear Medicine and Biology}. \textbf{96-97}:41-9.
	\item[] Fahey, F H, Kinahan, P E, Doot, R K, Kocak, M, Thurston, H and Poussaint, T Y. 2010. Variability in PET quantitation within a multicenter consortium. \textit{Medical Physics} \textbf{37(7)}:3660-6.
	\item[] Foley, R. 2012. \textit{When is true belief knowledge?} Princeton, New Jersey: Princeton University Press.
	\item[] Hachem, M, Tiberi, M, Ismail, B, Hunter, C R, Arksey, N, Hadizad, T, et al. 2016. \textit{The Journal of Nuclear Medicine}. \textbf{57(10)}:1612-7.
	\item[] Hatt, M, Cheze Le Rest, C, Albarghach, N, Pradier, O and Visvikis, D. 2011. PET functional volume delineation: a robustness and repeatability study. \textit{European Journal of Nuclear Medicine and Molecular Imaging}. \textbf{38}:663-672.
	\item[] Hatt, M, Lee, J A, Schmidtlein, C R, El Naqa, I, Caldwell, C, De Bernardi, E et al. 2017. Classification and evaluation strategies of auto-segmentation approaches for PET: Report of AAPM task group No. 211. \textit{Medical Physics}. \textbf{44(6)}:42.
	\item[] Higson, E, Handley, W, Hobson, M and Lasenby, A. 2017. Dynamic nested sampling: an improved algorithm for parameter estimation and evidence calculation. \textit{Statistics and Computing} \textbf{29. 5}: 891-913.
	\item[] Hofheinz, F, P\"otzsch, C, Oehme, L, Beuthien-Baumann, B, Steinbach, J, Kotzerke, J et al. 2012. Automatic volume delineation in oncological PET. \textit{Nuklearmedizin}. \textbf{51(01)}: 9-16.
	\item[] Jadvar, H and Parker, J A. 2005. \textit{Clinical PET and PET/CT}. London, UK: Springer Verlag.
	\item[] Kirov, A S, Schmidtlein, C R, Kang and H, Lee, N. 2012. 'Rationale, Instrumental Accuracy, and Challenges of PET Quantification for Tumor Segmentation in Radiation Treatment Planning', in Hsieh, C.-H. (ed.) Positron Emission Tomography-Current Clinical and Research Aspects. London, UK: Intechopen.
	\item[] Lee, J A. 2010. Segmentation of positron emission tomography images: Some recommendations for target delineation in radiation oncology. \textit{Radiotherapy and Oncology}. \textbf{96}: 302-7.
	\item[] Maguire, R P, eds, Leenders, K L, eds, Carson, R E, Cunningham, V C, Gunn, R N, van den Hoff, J et al. 2003. \textit{PET Pharmacokinetic Course}. Montreal, Qc: Montreal Neurological Institute.
	\item[] Mason, D L et al. 2022. Pydicom: An open source DICOM library. Available at: https://github.com/pydicom/pydicom (Accessed 15-07-2022).
	\item[] Meikle, S R, Sossi, V, Roncali, E, Cherry, S R, Banati, R, Mankoff, D et al. 2021. Quantitative PET in the 2020s: A Roadmap. \textit{Physics in Medicine and Biology}. \textbf{66}:56.
	\item[] Popper, K R. 2002. \textit{The Logic of Scientific Discovery}. London, UK: Routledge.
	\item[] Skilling, J. 2006. Nested sampling for general Bayesian computation. \textit{Bayesian Analyses}. \textbf{1(4)}: 833-59.
	\item[] Soret, M, Bacharach, S L and Buvat, I. 2007. Partial-volume effet in PET tumor imaging. \textit{Journal of Nuclear Medicine}. \textbf{48}:932-45.
	\item[] Speagle, J S. DYNESTY: a dynamic nested sampling package for estimating Bayesian posteriors and evidences. 2020. \textit{Monthly Notices of the Royal Astronomical Society}, \textbf{493(3)}: 3132-58.
	\item[] Tamal, M. A phantom study to assess the reproducibility, robustnes and accuracy of PET image segmentation methods against statistical fluctuations. 2019. \textit{PLoS One}. \textbf{14(7)}:18.
	\item[] Weeks, A.R. 1996. \textit{Fundamentals of Electronic Image Processing}. Bellingham, WA: SPIE Optical Engineering Press.
\end{harvard}
\end{document}